\documentclass[preprint,10pt]{elsarticle}

 \usepackage{graphics}
 \usepackage{graphicx}
 \usepackage{epsfig}

\usepackage{amssymb}

\journal{Physica A}

\begin{document}

\begin{frontmatter}



\title{Normal and anomalous diffusion of Brownian particles on disordered potentials}


\author{R. Salgado-Garc\'{\i}a}
\ead{raulsg@uaem.mx}
\address{Centro de Investigaci\'on en Ciencias, Universidad Aut\'onoma del Estado de Morelos. Avenida Universidad 1001, colonia Chamilpa, C.P. 62209, Cuernavaca Morelos, Mexico.}

\begin{abstract}
In this work we study the transition from normal to anomalous diffusion of Brownian particles on disordered potentials. The potential model consists of a series of ``potential hills'' (defined on unit cell of constant length) whose heights are chosen randomly from a given distribution. We calculate the exact expression for the diffusion coefficient in the case of uncorrelated potentials for arbitrary distributions. We particularly show that when the potential heights have a Gaussian distribution (with zero mean and a finite variance) the diffusion of the particles is always normal. In contrast when the distribution of the potential heights are exponentially distributed we show that the diffusion coefficient vanishes when system is placed below a critical temperature. We calculate analytically the diffusion exponent for the anomalous (subdiffusive) phase by using the so-called ``random trap model''. We test our predictions by means of Langevin simulations obtaining good agreement within the accuracy of our numerical calculations.

\end{abstract}


\end{frontmatter}



\section{Introduction}
\label{sec:introduction}

Transport of classical overdamped particles in random media has been intensively studied due to its relevance in several physical systems~\cite{bouchaud1990anomalous}. Due to its importance, several kind of models of transport in disordered media has been introduced in order to understand the underlying mechanisms originating the observed phenomenology~\cite{bouchaud1990anomalous,bouchaud1990classical}. Depending on the model type, the system can exhibit normal or anomalous transport accompanied  of normal or anomalous diffusion. For example, recent works have revealed that a system of overdamped particles on a random potential the unbiased diffusion is always normal independently of the correlations~\cite{goychuk2014anomalous,salgado2015unbiased}. In contrast, in models with uncorrelated Gaussian random force field exhibit anomalous diffusion when the system is unbiased (i.e., when the system is not subjected to a constant driving force and consequently it does not exhibit net transport)~\cite{bouchaud1990classical,sinai1983limiting,golosov1984localization}. Particularly, some attention has been addressed to models which exhibit a transition from normal to anomalous diffusion as a function of the temperature. For example, a simplified model exhibiting this feature is the so called random trap model~\cite{bouchaud1992weak}. In this model the phase space consist of a ``chain'' of traps, indexed by $i\in\mathbb{Z}$ with depth $\Delta E_i$. The time that a particle spent inside a trap has a distribution whose first moment diverges at a critical temperature $T_c>0$. This is a result that follows from the statistical  properties of the energy depths $\Delta E_i$, whose distribution is exponential. On the other hand, it has also proposed a model of Brownian particles with continuous disorder (a Gaussian squared potential) in which a dynamical transition also occurs at a finite temperature~\cite{touya2007dynamical}. 

In this work we study the particle-polymer model, a model which is characterized by having continuous dynamics but discrete disorder~\cite{salgado2013normal,salgado2014effective}. We will show that this type of models, which can be considered in the middle between the random trap models (which are discrete) and models with gaussian (continuous) disorder, are able to exhibit a transition from anomalous to normal diffusion as a function of the temperature. The  particle-polymer model consists of an ensemble of Brownian particles interacting with a random polymer. The polymer is built  up by concatenating at random (by means of some stochastic process) monomers of constant length which can be taken from a finite or infinite set of monomer types. The interaction of a given particle with the monomers defines the potential profile that the particle sees on a specific unit cell.  When the interaction is short-ranged we can think of this model as an ensemble of particles moving on a series of tracks, each track having a fixed random potential profile. Based on the theory developed in Ref.~\cite{salgado2014effective}, we will show that when the potential profile has barrier heights with normal distribution, the diffusion is always normal. In contrast, when the distribution of the potential barrier height is exponential, we will show that there is a finite temperature $T_c$ at which the system transits from normal to anomalous diffusion. We use the random trap model to approximate the behavior of the system in order to obtain an an analytical expression for the diffusion exponent in the anomalous  (subdiffusive) phase. 

In consequence this work is organized as follows. In section~\ref{sec:model} we introduce the model that we study and review some results already known for this system. In Section~\ref{sec:diffusion-coefficient} we calculate the diffusion coefficient for the systems by means of the Einstein relation which is known to be valid for our model. We also calculate explicitly  the diffusion coefficient for the cases in which the potential height has a Gaussian and exponential distributions respectively we show that in the latter case the diffusion coefficient is zero below a critical temperature. In Section~\ref{sec:anomalous} we study the case in which the potential heights are exponentially distributed and the temperature is below the critical one. Since in this case the diffusion coefficient is zero the system exhibit anomalous subdiffusion. We implement what we call the ``random trap'' approximation. This approximation is based in identifying the waiting time distribution for the random trap model with the distribution of the mean first passage time induced by the disorder.  This method allows us to obtain a good approximation for the diffusion exponent for the subdiffusive anomalous phase. We compare all the predictions made for this system against numerical simulations.  Finally in Section~\ref{sec:conclusions} we give the main conclusions of our work.

\section{The model}
\label{sec:model}

The particle-polymer model for transport in disordered media has been put forward in Ref.~\cite{salgado2013normal}. This model consists of an ensemble of non-interacting overdamped particles sliding over a 1D substrate. The substrate is model as a series of unit cells of constant length,  called monomers if the substrate is interpreted as a polymer, with which the particles are interacting. This model mimics in some sense, the motion of a given protein along a biological ``disordered'' polymer such as the DNA. The motion of a given particle along the referred substrate is ruled the stochastic differential equation,
\begin{equation}
\label{eq:model}
\gamma dX_t = \left( f(X_t) + F \right) dt  + \varrho_0 dW_t.
\end{equation}
Here $X_t$ represents the particle position at a time $t$, $f(x)$ is minus the gradient of the potential $V(x)$, which is the potential induced by the particle-polymer interaction and $W_t$ is a standard Wiener process.  The constants $\varrho_0^2$, $F$ and $\gamma $ are the noise intensity, the external driving force, and the friction coefficient respectively. According to  the fluctuation-dissipation theorem $\varrho_0^2 = 2 \gamma \beta^{-1}$, where $\beta$, as usual, stands for the inverse temperature times the Boltzmann constant, $\beta = 1/k_B T$.

Of course, the interaction of the particle with the polymer specifies the potential profile seen by the particle when it is located at a specific monomer.  If the particle-polymer interaction is short-ranged we can assume that the potential profile seen by the particle at a given monomer is defined exclusively by the monomer type at which the particle is located. Let us denote by            $\mathbf{a} := (\dots, a_{-1}, a_0, a_1, \dots)$ a specific realization of the random polymer, where $a_j$ is the monomer type at the $j$th unit cell. The variable $a_j$ can be interpreted as a random variable defining the interaction with the particle or defining the properties of the potential profile if the interaction is short-ranged. Thus, in the latter case, we can think of the potential $V(x)$ as a function of both, the particle position $x$ and the random variable $a_n$, where $n$ is index identifying the unit cell at which the particle is located. If the particle is located at the $n$th unit cell, then its position can be written as $x = nL + y$, where $L$ is the length of the unit cell and $y$ is the relative position of the particle on the monomer. Thus, the potential $V(x)$ can be written as a function $\psi$ of two  variables: the relative position $y $ and the random monomer type, $V(x) = \psi(y,a_n)$.  This model for disordered media was first studied in absence of noise in Ref.~\cite{salgado2013normal} and the authors showed  that the particle current and the effective diffusion coefficient can be known exactly for a large class of disordered potentials. Later on, in Ref.~\cite{salgado2014effective} it was shown that the particle current and diffusion coefficient can still be exactly in terms of quadratures. However in Ref.~\cite{salgado2014effective} the author was mainly interested in the case of biased transport, and, in fact, no formula for the diffusion coefficient in absence of driven force was done. Here we will  explore the unbiased diffusion based on the findings of Ref.~\cite{salgado2014effective}.

First of all we should remind that in Ref.~\cite{salgado2014effective} the author gave an exact expression for the first passage time (FPT) on a unit cell, averaged with respect to the noise, given by
\begin{equation}
T_1(\mathbf{a}) = \gamma \beta \sum_{m=1}^\infty e^{-m\beta F L} q_{+}(\mathbf{a})q_{-}[\sigma^{-m}(\mathbf{a})] 
+ \gamma \beta I_0(\mathbf{a}).
\label{eq:T1}
\end{equation}
In the above equation $\mathbf{a}$ stands for the realization of the polymer and $q_\pm$ and $I_0$ are integrals defined as
\begin{eqnarray}
\label{eq:def-qpm}
q_\pm (\mathbf{a} ) &:=& \int_0^L dx \exp\big( \pm \beta [ \psi(x,\mathbf{a}) - xF] \big),
\\
I_0 (\mathbf{a}) &=&  \int_0^L Q_-(x,\mathbf{a}) B_+(x,\mathbf{a})dx,
\label{I0}
\end{eqnarray}
where the functions $B_{\pm} : \mathbb{R} \times\mathcal{A}^\mathbb{Z} \to \mathbb{R} $ and $Q_{\pm} :  \mathbb{R} \times\mathcal{A}^\mathbb{Z} \to \mathbb{R} $ were defined as,
\begin{eqnarray}
\label{eq:def-Qpm}
Q_\pm (x,\mathbf{a}) &:=& \int_0^x dy \exp\big( \pm \beta [ \psi(y,\mathbf{a}) - yF] \big),
\\
\label{eq_def-Bpm}
B_\pm (x,\mathbf{a}) &:=&\exp\big( \pm \beta [ \psi(x,\mathbf{a}) - xF] \big).
\end{eqnarray}

The above expressions are valid for a large verity of disordered potentials, since these expressions were derived under some mild general assumptions. Specifically, the two restrictions imposed were $i)$ the \emph{ergodicity} of the random polymer when the sequence of monomer is seen as a discrete-time stochastic process and $ii$) the validity of the central limit theorem in its general form~\cite{gnedenko1968limit,gouezel2004central,chazottes2012fluctuations}. These two properties can be exploited in order to consider more general random potentials, particularly those that can give rise anomalous diffusion in absence of external driving forces. First, let us remind that the particle current can be written in terms of the FPT, averaged with respect to the noise and the polymer ensemble, as~\cite{salgado2014effective},
\begin{eqnarray}
\label{eq:drift}
J_{\mathrm{eff}}(F) &=& \frac{L}{\langle T_1 \rangle_{\mathrm{p}}},
\end{eqnarray}
where we used the notation $\langle \cdot  \rangle_{\mathrm{p}}$  for the expected value with respect to the polymer ensemble.
Particularly, when the Brownian particle has a short-range interaction with the polymer, we can assume that the potential only depends on one ``coordinate'' of the polymer (as we explained above)  and that the monomers along the polymer does not have correlations, i.e., the random polymer can be considered as a result of a realization of an infinite set of independent and identically distributed (i.i.d.) random variables. These hypotheses  allow write down a more explicit expression for $\langle T_1 \rangle_{\mathrm{p}}$ as follows,
\begin{eqnarray}
\langle T_1 ({a}) \rangle_{\mathrm{p}} &=& \gamma \beta \frac{ e^{-\beta F L}}{1- e^{-\beta F L}} \langle q_{+}({a})\rangle_{\mathrm{p}} \langle q_-(a) \rangle_{\mathrm{p}}  
+ \gamma \beta \left \langle I_0(a)\right \rangle_{\mathrm{p}}.
\label{eq:<T1>_uncorrelated}
\end{eqnarray}
Notice that in the last relation we have no longer made explicit the dependence of $T_1$ with the whole polymer, since the potential depends on only one monomer, and, due to translational invariance, the position of such a monomer does no longer matters. Thus,  the polymer average will only depend on the random variable $a$ and its distribution.

\section{The diffusion coefficient}
\label{sec:diffusion-coefficient}

It is known that in the class of disordered systems we are considering the Einstein relation remains valid~\cite{bouchaud1990anomalous}. This allows us to obtain an expression for the diffusion coefficient in absence of driving force.  If we denote by $D_{\mathrm{eff}}$  the effective diffusion coefficient for the system in absence of external driving force (i.e., for $F=0$), the Einstein relation states that
\[
D_{\mathrm{eff}} =  \frac{1}{\beta}\lim_{F\to 0}\frac{J_{\mathrm{eff}} (F)}{F}.
\]
To obtain an expression for the diffusivity in this case, first notice that $T_1$ diverges as $F$ goes to zero. It is easy to see that the asymptotic behavior is  giving by,
\[
\langle T_1 ({a}) \rangle_{\mathrm{p}}   \approx   \frac{\gamma}{F L}   Z(\beta) Z(-\beta)  
+ \gamma \beta \left \langle I_0(a)\right \rangle_{\mathrm{p}},
\]
where $Z (\beta)$ is defined as,
\[
Z(\beta) := \left \langle \int_0^L dx \exp\big( \beta \psi(x,{a}) \big) \right \rangle_{\mathrm{p}}.
\]
From the above expression we can see that the diffusion coefficient can be written straightforwardly as follows,
\[
D_{\mathrm{eff}} = \frac{L^2 }{\gamma \beta Z(\beta) Z(-\beta)},
\]
which is the effective diffusion coefficient at zero driving force, $F= 0$. Below we will use this expression to exactly calculate the diffusion coefficient in a specific potential model.

\subsection{The potential model}

\begin{figure}[h]
\begin{center}
\scalebox{0.35}{\includegraphics{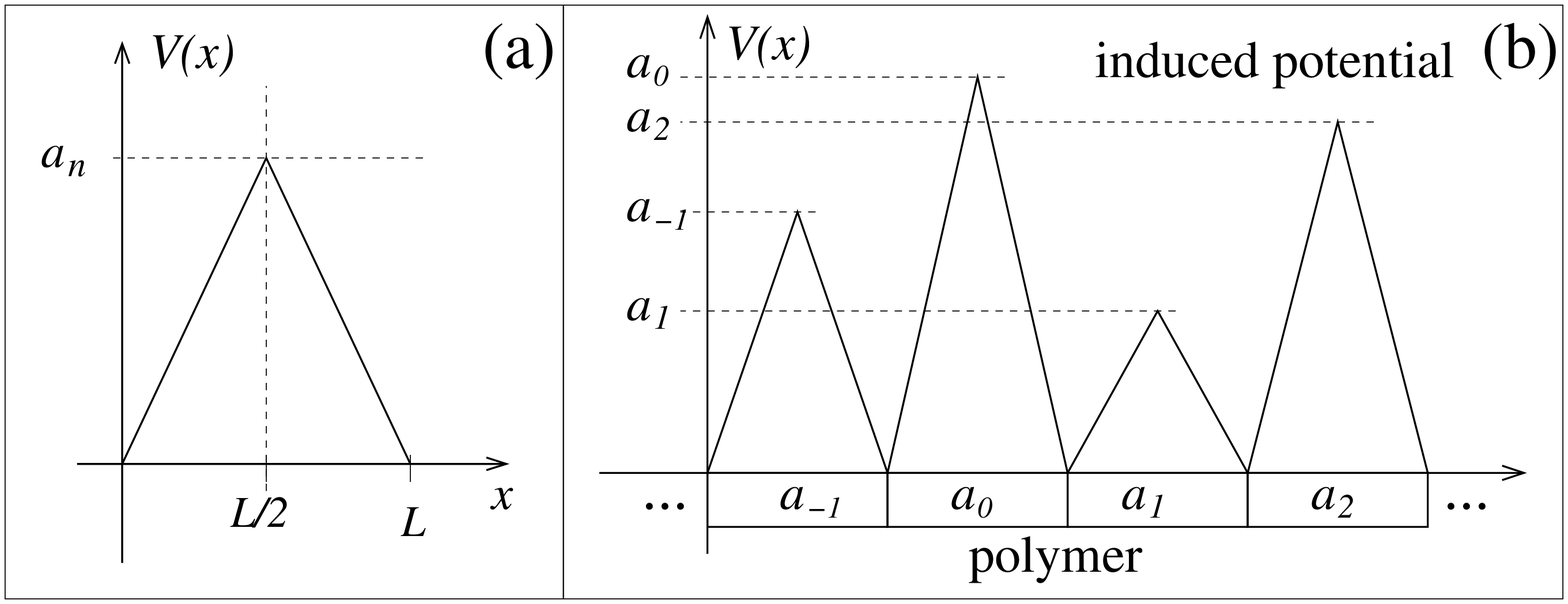}}
\end{center}
     \caption{
     Schematic representation of the potential model. (a) The potential profile on the $0$th unit cell. (b) A realization of the random potential with a few unit cells. The monomer type is labeled by $a_n$, with $n\in\mathbb{Z}$ and this label is in turn a random variable giving the height of the potential hill. There are as many monomer types as the number of values that can take any random variable $a_n$.
              }
\label{fig:potential}
\end{figure}

The model potential that we will use consists of a piecewise linear function. This makes the force field to be piecewise constant. Let $x$ be the particle position and let $n$ be the cell at which this particle is located. It is clear that $x$ can be written $x  = y +nL$, where $y$ is the relative position of the particle on the $n$th unit cell. Thus, the potential that the particle feels at $x$ is defined as,
\begin{equation}
V(x)= \psi (y,a_n) =  \left\{ \begin{array} 
            {r@{\quad \mbox{ if } \quad}l} 
   2a_n y/L   &  0\leq  y<L/2    \\ 
   2a_n (L-y )/L  &  L/2 \leq y < L.       \\ 
             \end{array} \right. 
\label{eq:potential-model}
\end{equation}
where $\mathbf{a}:=(\dots, a_{-1},a_0, a_1,\dots)$ represents a realization of the random polymer. Here $a_n$ stands for the potential height and this quantity specifies a polymer type (see Fig.~\ref{fig:potential} for a schematic representation of the potential model). Thus the set $\{\dots, a_{-1},a_0, a_1,\dots\}$  can be considered as a realization of an infinite set of i.i.d.~random variables. The reason to take these random variables as independent is because the polymer is assumed to be uncorrelated. On the other hand, the reason to assume that the random variables are identically distributed is just to fulfill the condition of translational invariance. 
%

Let $a$ be a random variable with the same distribution of any $a_n$ for $n\in \mathbb{Z}$. If the random variable $a$ is assumed to be discrete taking  values on a finite set $\{ V_j : 0\leq j\leq N\}$, then, the number of monomer types is finite. However, the random variable $a$ is not restricted to take values in a finite set nor a discrete one. The random variable can even take values in a continuous set and the hypotheses of polymer ergodicity and translational invariance is not violated. In this case the number of monomer types will be uncountably infinite. It is particularly interesting to investigate the case in which the set of possible potential heights is $\mathbb{R}$ or $\mathbb{R}^+$, i.e., when the height distribution have unbounded support.  In this case, we are in a situation in which the potential hills have unbounded height and the particle could stay long times in a valley along the polymer. In this situations we would expect to have a competition between the thermal activation, taking the particles out of the valley, and the height of potential barriers, slowing down the dynamics of the particles. Below we show that to slow down the diffusivity is such a way that the diffusion coefficient be zero it is not enough that the height distribution have unbounded support. Some additional features on the distribution are needed in order to have a zero diffusion coefficient. In particular we show that when the height distribution is gaussian, with zero mean and finite variance, the diffusion coefficient never vanishes. On the other hand, when the height distribution is exponential, the diffusion coefficient can be zero below a critical temperature, i.e., the system ``freezes'' at a finite (positive) temperature.

\subsection{Gaussian heights}

Let us analyze the case in which the height distribution is Gaussian with zero mean and variance $\sigma^2$. The variance $\sigma^2$  is an intrinsic parameter characterizing the disorder. Along this line we can say that the probability density function $\rho (a)$ of the height of the potential hills along the polymer is given by,
\[
\rho_{\mathrm{g}} (a) = \frac{1}{\sqrt{2\pi \sigma^2} }\exp{\left( -a^2/2\sigma^2\right)}.
\]
Note that the distribution allows negative values for the potential ``height'' which mean that a typical realization of the potential contains not only ``potential hills'' but also ``potential wells''. 

In order to calculate the exact diffusion coefficient we should first evaluate $Z(\beta)$, which can be written as 
\[
Z (\beta) := \left \langle \int_0^L dx \exp\big(  \beta \psi(x,{a}) \big) \right \rangle_{\mathrm{p}} =
\int_{-\infty}^\infty da \int_0^L dx \exp\big[\beta \psi(x,{a}) \big] \rho_{\mathrm{g}} (a).
\]
First notice that the potential $\psi (y,a_n)$ can be factorized as the random height $a_n$ times a dimensionless function $\varphi$ which is piecewise linear,  i.e., $ \psi (y,a_n) = a_n \varphi(y)$, where
\begin{equation}
\label{eq:def:phi}
 \varphi (y) =  \left\{ \begin{array} 
            {r@{\quad \mbox{ if } \quad}l} 
   2 y/L   &  0\leq  y<L/2    \\ 
   2 (L-y )/L  &  L/2 \leq y < L.       \\ 
             \end{array} \right. 
\label{eq:potential-model}
\end{equation}
Then it is easy to see that the integral $Z(\beta)$ can be written as , 
\begin{eqnarray}
Z(\beta) 
&=&\int_0^L dx  \int_{-\infty}^\infty da  \exp\big[\beta a \varphi(x) \big] \frac{1}{\sqrt{2\pi \sigma^2} }\exp{\left( -a^2/2\sigma^2\right)},
\nonumber
\end{eqnarray}
where the integral with respect to the variable $a$ can be easily evaluated to give,
\[
Z(\beta) = \int_0^L dx  \exp\left[ \beta^2\sigma^2 \varphi^2(x)/2\right].
\]
Now, the above result  allows us to calculate $Z(\beta)$ for our model explicitly in terms of quadratures. After a few manipulations it is possible to see that, 
\[
Z(\beta) = \frac{L}{\beta \sigma }G(\beta \sigma), 
\]
where we have defined $G$ as the function,
\[
G(\beta \sigma) = \int_{0}^{\beta \sigma} 	\exp{\left(x^2/2\right)}dx.
\]
The result for $Z(\beta)$ allows us to write the diffusion coefficient as
\begin{equation}
\label{eq:Deff_gaussian}
D_{\mathrm{eff}} = \frac{\beta \sigma^2}{\gamma G^2(\beta \sigma)  }.
\end{equation}

Notice that when the temperature is large, i.e., when $\beta \sigma \ll 1$, we can approximate the function $G(\beta \sigma)$ as,
\[
G(\beta \sigma ) \approx \beta \sigma e^{\beta^2 \sigma^2/2} \qquad \mbox{for } \ \beta \sigma \ll 1.
\]
Then, the diffusion coefficient take the form,
\begin{equation}
D_{\mathrm{eff}} \approx \frac{e^{-\beta^2 \sigma^2} }{\gamma\beta    } \qquad \mbox{for } \ \beta \sigma \ll 1,
\end{equation}
which is referred to as the super-Arrenious factor as in the case of continuously disordered Gaussian potentials~\cite{zwanzig1988diffusion,dean2007effective,touya2007dynamical}. 
\begin{figure}[h]
\begin{center}
\scalebox{0.4}{\includegraphics{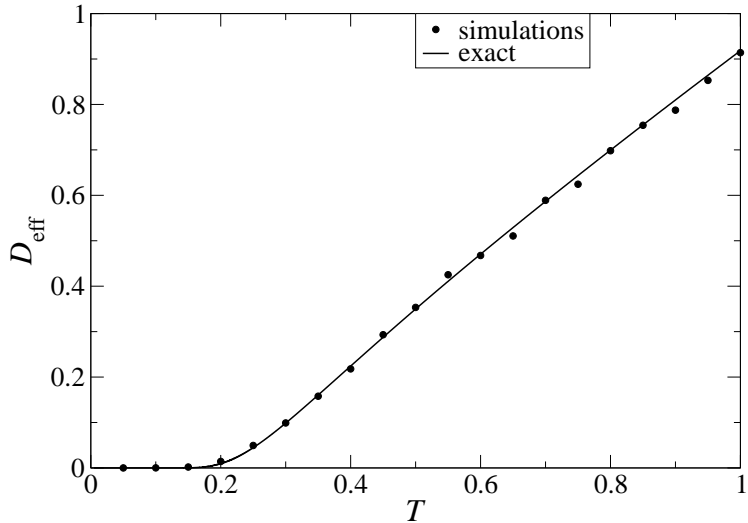}}
\end{center}
     \caption{
     Effective diffusion coefficient for the random potential with Gaussian height distribution. We display the diffusion coefficient given by the exact expression of Eq.~(\ref{eq:Deff_gaussian}) (solid line) and by the numerical experiment (filled circles). For the numerical simulation we generate five polymer of $2\times 10^6$ monomers. On each monomer we assumed that the potential profile has the form of a ``hill'', which is given by Eq.~(\ref{eq:potential-model}). The potential heights $\{a_n\}$ were defined by generating a series of $2\times10^6$ normally distributed random number with zero mean and variance $\sigma^2=1$. Then we placed $4\,000$ particles along the polymer which were subjected to the Langevin dynamics simulations. The obtained trajectories were used to estimate the mean square displacement, which in turns gave us an estimation for the effective diffusion coefficient.
             }
\label{fig:gaussian}
\end{figure}
%

In Fig.~\ref{fig:gaussian} we display the effective diffusion coefficient analytically calculated using Eq.~(\ref{eq:Deff_gaussian}). In the same figure we also display the effective diffusion coefficient obtained from the numerical simulation of an ensemble of overdamped particles in a Gaussian random potential. The random potential was obtained by generating five series of $2\times 10^6$ Gaussian random numbers with zero mean and variance one ($\sigma^2 =1$). Then, we used these random numbers as the heights of the potential hill comprising the polymer. Thus every polymer has $2\times 10^6$ monomers. Next, we placed $4\,000$ particles equally spaced on every polymer and such particles were subjected to the Langevin dynamics during a total time of $10^5$ arb. units. We fixed the friction constant $\gamma$ equal to one. Then we used the resulting trajectories to estimate the mean square displacement, from which we obtained the corresponding estimation of the diffusivity. As we can see in Fig.~\ref{fig:gaussian} w have a good agreement (within the accuracy of our simulations) between the exact curves an the  numerical experiment.

\subsection{Exponential heights}

We now consider the case in which the height distribution is exponential. Notice that for the exponential distribution there is no negative heights, which means that there is no potential ``wells'' but only potential ``hills''. We could think that, since there is no potential wells the probability that a given particle gets stuck is, in this case, lower than in the case of the Gaussian height distribution. This is because in the present situation the particle has to overcome only the potential barrier due to the hills, contrary to the Gaussian case in which the particle has to overcome the potential barrier when it is located at a well (negative height) plus the potential barrier due to the presence of a hill (positive height). However we will show that, contrary to our intuitive arguments, the diffusivity in the exponential case is \emph{lower} than in the Gaussian case. This slowing down of the diffusivity is still much more dramatic, since at sufficiently low temperatures the diffusion coefficient vanishes. This means that there occur a dynamical transition from normal to anomalous subdiffusion as the temperature decreases.  Moreover, we will also show that the mean square displacement of the Brownian particles grows in time asymptotically as $\Delta X_t^2 \sim t^\alpha$ where $\alpha$, the diffusion exponent, can be analytically calculated by using the approximation provided by the random trap model for subdiffusion.

First let us calculate the diffusion coefficient for this model. We assumed that the height distribution is exponential, which means that the corresponding probability density function reads as,
\[
\rho_{\mathrm{e}}(a) = \frac{\exp\left(-a/\lambda\right) }{\lambda}.
\]
As in the Gaussian case we need to calculate the function $Z(\beta)$ which in this case can be written as,
\begin{eqnarray}
Z(\beta) &=& \int_{0}^\infty da \int_0^L dx \exp\big[\beta \psi(x,{a}) \big] \rho_{\mathrm{e}} (a)
\nonumber
\\
&=&\int_0^L dx  \int_{0}^\infty da  \exp\big[\beta a \varphi(x) \big] \frac{\exp\left(-a/\lambda\right) }{\lambda} .
\nonumber
\\
&=&\int_0^L  \frac{dx}{1-\beta \lambda \varphi(x)}
\nonumber
\end{eqnarray}
Notice that the last integral diverges when $ \beta \lambda \varphi(x) = 1$ at some $x\in [0,L]$. This is of course possible for temperatures low enough. Indeed if we recall the definition of $\varphi$ [see Eq.~(\ref{eq:def:phi})] we observe that $0\leq \varphi(x) \leq 1$ for all $x\in[0,L]$ and in particular the upper bound is reached at $x=L/2$. This implies that when the critical (inverse) temperature is given by $\beta_\mathrm{c} = 1/\lambda$. If $\beta \geq \beta_\mathrm{c}$ then $Z{(\beta)}$ diverges and consequently the diffusion coefficient is zero. Conversely, if $\beta < \beta_\mathrm{c}$ then $Z(\beta)$ is finite and the diffusion coefficient is positive. In the latter case the integral can be done exactly for our potential model, giving,
\[
Z(\beta)  = -\frac{L}{ \beta \lambda} \ln (1-\beta \lambda) \qquad \mbox{for} \quad \beta < \beta_{\mathrm{c}} = 1/\lambda.
\]
The above result allows us to write down explicitly the effective diffusion coefficient, 
\begin{equation}
\label{eq:Deff-exponential}
D_{\mathrm{eff}} = \frac{\beta \lambda^2}{ \gamma} \frac{1}{| \ln(1-\beta \lambda)\ln(1+ \beta \lambda)|} \qquad \mbox{for} \quad \beta < \beta_{\mathrm{c}} = 1/\lambda.
\end{equation}

To perform the simulations of the Langevin dynamics as well as to numerically evaluate $D_{\mathrm{eff}}$ let us introduce dimensionless  quantities. First notice that $\beta \lambda$ is a dimensionless parameter which allows us to define naturally a dimensionless temperature $\tilde T$ as
\[
\tilde T := \frac{1}{\beta \lambda}.
\]
Similarly we should observe that $\lambda/\gamma$ has the same units as the diffusion coefficient. This suggest to define a dimensionless effective diffusion coefficient $\tilde D_{\mathrm{eff}}$ as 
\[
\tilde D_{\mathrm{eff}} :=  \frac{\gamma D_{\mathrm{eff}} }{\lambda}.
\]
Using the above definition we can rewrite the expression for the diffusion coefficient in terms of dimensionless parameters,
\[
\tilde D_{\mathrm{eff}} = \frac{1}{\tilde T } \frac{1}{\left| \ln\left(1-\frac{1}{\tilde T }\right)\ln\left(1+\frac{1}{\tilde T }\right)\right|} \qquad \mbox{for} \quad \tilde T > \tilde T_{\mathrm{c}} = 1.
\]
\begin{figure}[h]
\begin{center}
\scalebox{0.4}{\includegraphics{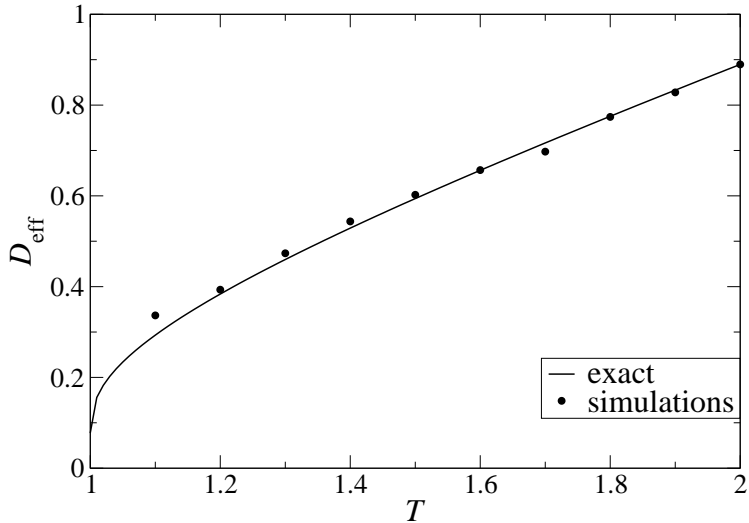}}
\end{center}
     \caption{
    Effective diffusion coefficient for the random potential with exponential height distribution. We display the diffusion coefficient given by the exact expression of Eq.~(\ref{eq:Deff-exponential}) in terms of dimensionless quantities (solid line) and by the numerical experiment (filled circles). For the numerical simulation we generated five polymer of $2\times 10^6$ monomers. On each monomer we assumed that the potential profile has the form of a ``hill'', which is given by Eq.~(\ref{eq:potential-model}). The potential heights $\{a_n\}$ were defined by generating a series of $2\times10^6$ exponentially distributed random number with zero mean and variance $\sigma^2=1$. Then we placed $4\,000$ particles along the polymer which were subjected to the Langevin dynamics simulations. The obtained trajectories were used to estimate the mean square displacement, which in turns gave us an estimation for the effective diffusion coefficient.
          }
\label{fig:exp_unbiased}
\end{figure}
%

In Fig.~\ref{fig:exp_unbiased} we display the effective diffusion coefficient above the critical temperature. To perform the numerical simulations we have chosen the parameters $\lambda  =  L = 1$. As we saw above we have that the dimensionless critical temperature $T_\mathrm{c}$ equals one. We observe that within the accuracy our simulations the numerically obtained diffusion coefficient agrees with the one obtained by means of the exact closed formula given in Eq.~(\ref{eq:Deff-exponential}).

\section{Anomalous subdiffusion: an analytical approach}
\label{sec:anomalous}

We have seen that when the heights are exponentially distributed the diffusion coefficient goes down to zero  for all the temperatures below a critical temperature $T_c = \lambda/k_B$. This fact implies that the mean square displacement $\Delta X_t^2$ grows in time lower than linear, i.e., 
\begin{equation}
\label{eq:diff-exponent}
\Delta X_t^2 \sim t^\alpha
\end{equation}
where $\alpha$ is what we call the \emph{diffusion exponent}. The diffusion exponent characterizes the subdiffusive regime, indicating in some sense how low is the diffusivity of the Brownian particles on the random potential. Here we will implement an analytical treatment in order to obtain an approximate  expression for the diffusion exponent $\alpha$. First of all we should notice that the origin of the subdiffusive behavior lies on the statistics of the potentials. The height of the potential barriers are responsible of the (long) time that a particle remains in a given cell. The higher potential barrier the larger time the particle spent on the cell.

Thus we need to estimate the distribution of the time that a particle takes to go from a given cell to the adjacent one. This time will be latter identified as a ``waiting time'' in a random trap model. To perform such an estimation we assume that the time that a particle need to go from a unit cell to the adjacent one is given by the inverse of the Kramers escape rate $\kappa_{\mathrm{esc}}$. This quantity is given by~\cite{hanggi1990reaction}
\[
\kappa_{\mathrm{esc}} = \frac{1}{2\pi} \sqrt{ V^{\prime\prime} (x_{\mathrm{min}})V^{\prime\prime} (x_{\mathrm{max}})} \exp\big(- \beta\big[V (x_{\mathrm{max}})-V (x_{\mathrm{min}}) \big] \big).
\]  

It is important to remark that the Kramers escape rate is an approximation to estimate the time that a particle need to jump to the next unit cell. This is because the ``true'' waiting time (let us call $T_w$) is a random variable that depends on the two underling stochastic processes present in the system: the ``Gaussian white noise'' (i.e., the Wiener process $W_t$) and the random polymer $\mathbf{a}$. We actually pretend that waiting time is nearly equal to  the inverse Kramers $\tau: = \kappa_{\mathrm{esc}}^{-1}$ escape rate, which is asymptotically equal only in the limit of zero noise strength. The latter is due to the fact that $\kappa_{\mathrm{esc}}^{-1}$ is in fact the average of  $T_w$ with respect to the noise, i.e. $\tau = \langle T_w \rangle_{\mathrm{p}}$. Thus we naturally  have  that $\tau$ is a random variable that only depends on the realization of the random polymer $ \mathbf{a}$. Moreover, since the potential model we have introduced only depends on one coordinate of the random polymer, we have that $\tau$ only depends on one random variable (or, in other  words, $\tau$ only depends on one coordinate or $\mathbf{a}$).

In this case it is easy to see that the escape rate is approximately given by,
\[
\kappa_{\mathrm{esc}} \approx \frac{\lambda}{2\pi L^2} e^{-\beta a},
\]
which means that the waiting time can be approximated as
\begin{equation}
\tau(a) =  \frac{2\pi L^2}{\lambda} e^{\beta a}.
\end{equation}
A direct calculation shows that the  distribution function for $\tau$ can be written as
\[
f_\tau (t) \approx \frac{1}{2\pi \beta} \left(\frac{\lambda t}{2\pi L^2} \right)^{-(\frac{1}{\beta\lambda} +1)},
\]
for $t$ large enough.

Then, following the assumption that the distribution of $\tau$ corresponds to the waiting time distribution of a random trap system, we have, according to the theory developed in Ref.~\cite{bouchaud1990anomalous}, that the diffusion exponent $\alpha$ defined through Eq.~(\ref{eq:diff-exponent}) is given by,
\begin{equation}
\alpha = \frac{2}{1+{\beta \lambda}}.
\end{equation}
In terms of the dimensionless quantities, defined above, $\alpha$ can be rewritten as,
\begin{equation}
\label{eq:alpha-difussivity}
\alpha = \frac{2}{1+\frac{1}{\tilde{T}}}.
\end{equation}
\begin{figure}[h]
\begin{center}
\scalebox{0.4}{\includegraphics{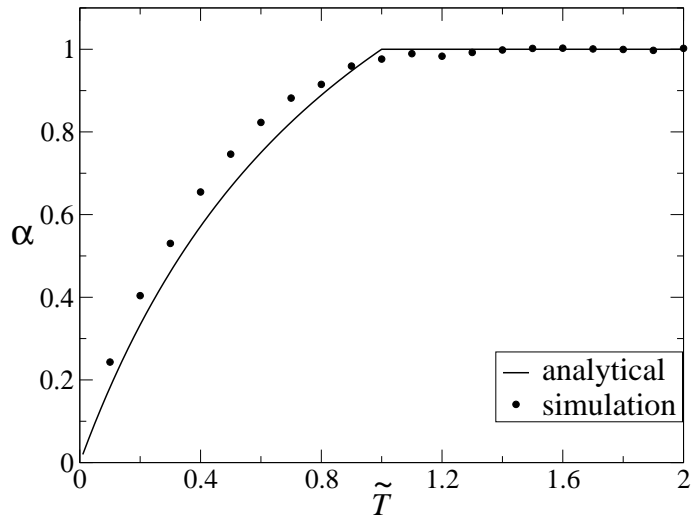}}
\end{center}
     \caption{
     Diffusion exponent for the random potential with exponential height distribution. The figure shows the diffusion exponent obtained from the approximate analytical expression of Eq.~(\ref{eq:alpha-difussivity}) (solid line) and from the numerical experiments (filled circles). The numerical experiment was done by simulating the stochastic dynamics of $20\,000$ particles, every particle on a different realization of the polymer. These trajectories were used to obtain the mean square displacement as a function of time. The latter is assumed to behave as $t^\alpha$ and from a fit of least squares we estimated the diffusion exponent $\alpha$.
          }
\label{fig:exp_unbiased-anomalous}
\end{figure}
%

In order to verify the approximations made to obtain the analytical expression for the diffusion exponent we have performed numerical simulations. Firstly we made ``random polymers'' by producing five series of $2\times 10^6$ exponentially distributed random numbers. Each series represents a random polymer  of $2\times 10^6$ symbols long. Then we placed $4\times 10^3$ particles evenly spaced along every polymer. Next the particles were subjected to the Langevin dynamics, from which we estimated the mean square displacement as a function of time. This procedure yields  the effective diffusion coefficient which is displayed in Fig.~\ref{fig:exp_unbiased-anomalous}. As we see, the behavior of the analytically calculated diffusion exponent is consistent with the simulations. This result allows us to say that the random trap model is a good approximation to the problem of an overdamped particles in a random exponential potential in presence of thermal noise.

\section{Discussion and conclusions}
\label{sec:conclusions}

In this work we have studied the particle-polymer model for diffusion on random potentials. We considered an ensemble of overdamped particles moving on a one dimensional random substrate which we interpret as a polymer. The substrate was assumed to interact with the particles in such a way  that the substrate can be thought as a series of tracks (unit cells) each track having a fixed potential profile. We assumed that the potential height on each track is a random variable and we explored the cases in which the potential heights have Gaussian distribution and exponential distribution. We calculated the exact diffusion coefficient in both cases and we found that when the height distribution is Gaussian the diffusion is normal at all temperatures. In contrast, when the potential heights are exponentially distributed we found that the diffusion coefficient vanishes for low enough temperatures. For the potential model with exponential height distribution we were able to exactly determine the critical temperature $T_{\mathrm{c}}$. For temperatures below $T_{\mathrm{c}}$ we assumed that the system can be seen as a random trap system and we estimated the waiting time distribution by using the Kramers escape rate. Within this approximation we were able to obtain the diffusion exponent as function of the temperature in a closed way. We performed  simulations of the corresponding  Langevin dynamics which confirmed our theoretical results within the accuracy of our numerical experiments.

\section*{Acknowledgments} 

The author thanks CONACyT for financial support through Grant No. CB-2012-01-183358.


%


\bibliographystyle{model1-num-names}
\bibliography{AnomDiff_UnboundedPot_ref}







\end{document}